\documentclass[aps,prl,twocolumn]{revtex4}
\usepackage[utf8]{inputenc}
\usepackage{amsmath}
\usepackage{commath}
\usepackage[pdftex]{graphicx}
\usepackage{siunitx}
\usepackage{hyperref}

\DeclareMathOperator{\atan}{atan}

\newcommand{\fet}[1]{\mathbf{#1}}
\newcommand{\vd}[2]{\frac{\delta #1}{\delta #2}}
\newcommand{\ex}{\text{ex}}

\newcommand{\ext}{\text{ext}}
\newcommand{\qEW}{\text{qEW}}
\newcommand{\fv}[1]{\left\langle #1 \right\rangle}

\begin{document}
\title{Depinning exponents of thin film domain walls depend on disorder strength}
\author{Audun Skaugen}
\author{Lasse Laurson}
\affiliation{Computational Physics Laboratory, Tampere University, P.O. Box 692, FI-33014 Tampere, Finland}

\begin{abstract}
Domain wall dynamics in ferromagnets is complicated by internal degrees of freedom of the domain walls. We develop a model of domain walls in disordered thin films with perpendicular magnetic anisotropy capturing such features, and use it to study the depinning transition. For weak disorder, excitations of the internal magnetization are rare, and the depinning transition takes on exponent values of the quenched Edwards-Wilkinson equation. Stronger disorder results in disorder-dependent exponents concurrently with nucleation of an increasing density of Bloch lines within the domain wall.
\end{abstract}

\maketitle

Domain walls (DWs) driven by applied magnetic fields in disordered
ferromagnets constitute a paradigmatic system exhibiting a depinning
transition between pinned and moving phases at vanishing temperatures $T$~\cite{zapperi1998dynamics,bustingorry2012thermal,pardo2017universal,
caballero2018magnetic} as well as slow thermally
activated creep motion for finite $T$~\cite{metaxas2007creep}. Related
phenomena include the Barkhausen effect~\cite{durin2006science,cizeau1997dynamics,laurson2014universality}, where scale-free jumps of DWs driven by a slowly changing external field are measurable as magnetic "crackling noise"~\cite{sethna2001crackling}. Thus, DWs are
often considered to belong to a broader class of driven systems displaying
similar phenomena, including also, e.g., cracks~\cite{laurson2013evolution}, 
contact lines~\cite{ertacs1994critical},
and grain boundaries~\cite{moretti2004depinning}.

A key class of %statistical physics 
models of such systems are driven elastic
interfaces in random media~\cite{narayan1993threshold} where one typically
assumes purely dissipative dynamics at $T$ small enough that creep can be ignored~\cite{zapperi1998dynamics}. %, i.e., local interface velocity proportional to the total force~\cite{zapperi1998dynamics}. 
Examples include simple models such as the quenched Edwards-Wilkinson (qEW) equation~\cite{edwards1982surface,kardar1986dynamic}. 
However, a crucial feature of magnetic DWs is that 
there are often significant non-dissipative effects related to the
magnetization direction inside the DW. This is most dramatically illustrated by the %phenomenon of 
Walker breakdown in 1D nanowires, where the internal magnetization begins to precess at a specific driving field magnitude, leading to a sharp drop in the DW propagation velocity~\cite{schryer1974motion}.
The depinning dynamics of point-like DWs with an internal degree of freedom in 1D systems can be dramatically changed by the Walker breakdown effect, leading to a series of transitions between a pinned and depinned DW as the driving field increases~\cite{lecomte2009depinning}.

For line-like DWs in 2D thin films with perpendicular magnetic anisotropy (PMA), instead of the internal in-plane magnetization rotating uniformly together, it can vary along the DW, resulting in formation of 1D domain wall-like structures known as Bloch lines (BLs) inside the DW~\cite{slonczewski1974theory,herranen2015domain,hutner2019multistep}. The motion of BLs, separating regions of different chiralities of the Bloch DW, mediates large-scale precession of the DW magnetization in an analogous manner to dislocation motion mediating plastic flow in crystals. Such effects were recently studied by full micromagnetic simulations of Barkhausen noise~\cite{herranen2019barkhausen}. However, micromagnetic simulations describing the magnetization dynamics everywhere in the system are limited to small system sizes,
%in that reachable DW lengths are restricted to a few micrometers, 
resulting in significant finite size effects. 

In this Letter, we study the depinning dynamics of thin film DWs in large systems (up to two orders of magnitude larger than in recent micromagnetic simulations~\cite{herranen2019barkhausen}) by developing a reduced model able to describe BL dynamics inside the DW while including only the degrees of freedom of the line-like DW itself. 
Strikingly, and contrary to what one observes in simple elastic line models of DWs neglecting the internal degrees of freedom, we find that the depinning exponents evolve with the disorder strength. We interpret this variation as a slow crossover from the universality class of the qEW equation -- describing DWs in weakly disordered films with a low BL 
density -- to another class in the limit of strong disorder. We argue that this crossover originates from the spatially heterogeneous dynamic arrangement of BLs affecting locally the DW mobility in the strong disorder regime. Our results thus reveal a previously unknown paradigm of disorder-dependent criticality at the depinning transition of DWs with internal degrees of freedom.

We formulate a %computationally efficient 
model of DWs in PMA films by viewing the Landau-Lifshitz-Gilbert (LLG) equation in terms of the polar angles $\theta$ and $\phi$ of the magnetization vector $\fet m = \cos \theta \fet e_z + \sin\theta(\cos\phi \fet e_x + \sin\phi \fet e_y)$
as a dissipative Euler-Lagrange equation, i.e.,
%\begin{align}
%  %\pd{}{t}
 $\frac{\partial}{\partial t}\left( \vd{\mathcal{L}}{\dot \theta} \right) - \vd{\mathcal{L}}{\theta} + \vd{F}{\dot \theta} = 0$ and % \nonumber \\
  %\pd{}{t}
  $\frac{\partial}{\partial t}\left( \vd{\mathcal{L}}{\dot \phi} \right) - \vd{\mathcal{L}}{\phi} + \vd{F}{\dot \phi} = 0.$ %\nonumber 
%\end{align}
Here the Lagrangian $\mathcal{L}[\theta,\phi] = T[\theta,\phi] - E[\theta,\phi]$ comprises of a “kinetic” part
$T[\theta,\phi] = \frac{M_s}{\gamma}\int \phi\dot\theta \sin\theta d\fet x$, where $M_s$ is the saturation magnetization and $\gamma$ the gyromagnetic ratio, and the energy functional $E[\theta,\phi] = \int \left[A_\ex (\nabla \theta^2 + \nabla\phi^2\sin^2\theta) - K_u\cos^2\theta - B_a M_s \cos\theta\right] d\fet x + E_d[\theta,\phi]$, using a local approximation for the demagnetization energy $E_d = - \frac 1 2 \mu_0 M_s^2 \int \left[N_n(\fet m \cdot \fet n)^2 + m_z^2\right]d\fet x$, where $\fet n$ is the unit vector perpendicular to the DW, and $N_n$ is given to lowest order in the film thickness $\Delta$ as $\frac{\Delta}{\pi D}\ln 2$ \cite{skaugen2019analytical}.
%\end{align}
The dissipation functional is given by
%\begin{align}
$F[\theta,\phi] = \frac{\alpha M_s}{2\gamma}\int \left( \dot \theta^2 + \dot\phi^2 \sin^2\theta \right)d\fet x$,
%\end{align}
where $\alpha$ is the Gilbert damping constant. We derive a local description by changing variables to coordinates co-moving with the DW, given by %\[
$\fet x = \fet r(s, t) + \rho \fet n(s, t)$, %\]
where $\fet r(s)$ is a parameterized curve describing the DW, $\fet n(s) = \ell^{-1}\fet e_z \times \fet u$ is the normal vector to the wall, $\fet u = \pd{\fet r}{s}$ is the tangent, 
$\ell = |\fet u|$ is the length of the tangent, and $\rho$ denotes the projected signed distance from $\fet x$ to $\fet r(s)$. Integrating over the normal coordinate $\rho$ and truncating to second order in the physical quantities, we find $\mathcal{L}$ and $F$ %a Lagrangian and dissipation functional 
for the quantities $\fet r$, $\phi$ and $D$ from which we can derive dynamical equations. We then choose the specific parameterization of a graph $\fet r(x, t) = x\fet e_x + h(x,t)\fet e_y$ and assume that the slope $\pd{h}{x}$ is small. Approximating the DW width as the constant $D = \sqrt{\frac{A_\ex}{K_u - \frac 1 2 \mu_0 M_s^2}}$, we find
\begin{align}
    \dot \phi + \alpha \frac{\dot h}{D} &= 2\frac{\gamma A_\ex}{M_s D}h'' - \gamma B_a, \label{eq:7} \\
    \alpha \dot \phi - \frac{\dot h}{D} &= 2\frac{\gamma A_\ex}{M_s}\phi'' - \frac{\gamma N_n}{2}\mu_0 M_s \sin[2(\phi-\chi)], \label{eq:8}
\end{align}
where primes denote differentiation wrt. $x$, and $\chi = \atan h'$ is the angle of the DW wrt. a flat, horizontal configuration. Notice that without internal degrees of freedom (i.e., for $\dot \phi=0$ and $\phi = \chi$), Eq.~(\ref{eq:7}) reduces to the qEW equation, while neglecting the spatial derivatives results in the "1D model" of DW dynamics for a constant %DW width
$D$~\cite{thiaville2006domain}.
The applied field $B_a$ includes quenched disorder modelled as a random out-of-plane magnetic field
$B_a(\fet r) = B_\ext + \eta(\fet r)$,
where the disorder is drawn from a normal distribution with mean 0 and standard deviation $\sigma$. This corresponds to random field disorder; random bond disorder is expected to result in the same critical behaviour~\cite{rosso2007numerical}. We ensure a spatial correlation length $\xi$ such that 
   $ \fv{\eta(\fet r)\eta(\fet r')} = \sigma^2 \exp[-\frac{|\fet r - \fet r'|^2}{\xi^2}]$
by multiplying an
uncorrelated array of random numbers in $k$ space with 
$\propto \exp(-k^2\xi^2/8)$, and Fourier-transforming back to real space. Linear interpolation is employed in the $h$ direction to compute the value of $\eta$ at a given point $h(x)$. %along the DW.

\begin{figure}
    \centering
    \includegraphics[width=0.5\textwidth]{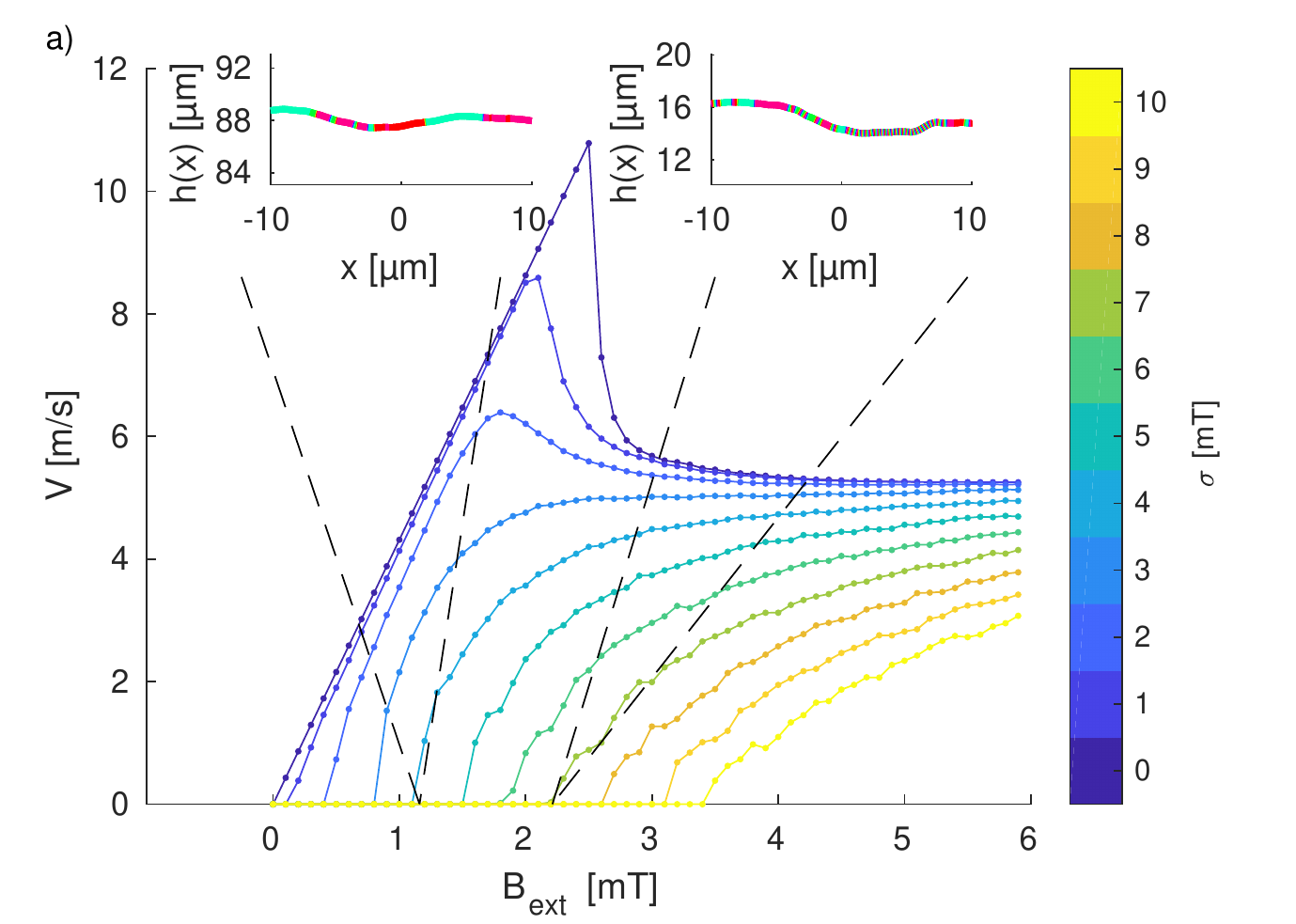}
    \includegraphics[width=0.5\textwidth]{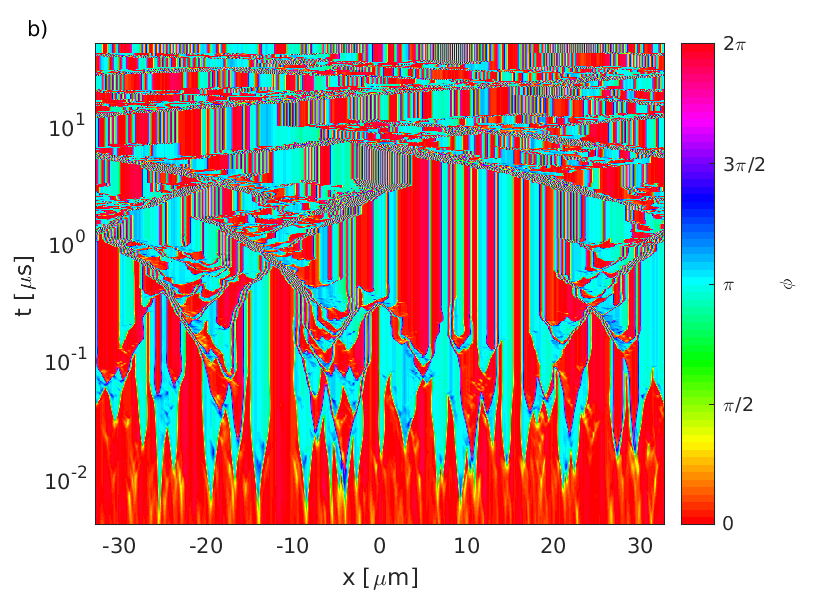}
    \caption{a) Steady-state DW velocity $V(B_\ext)$ 
    for different
    $\sigma$-values for $L \approx 65.5~\si{\micro\meter}$. Insets: 
    Snapshots of parts of the DW at 
    $B_\ext=B_c$ for $\sigma=4$ mT (left) and $\sigma=7$ mT (right),
    with color indicating the local $\phi$-value [colorbar in b)]. 
    b) Space-time map of $\phi$ at $B_\ext=B_c \approx 2.22~ \si{\milli\tesla}$ and $\sigma = 7~\si{\milli\tesla}$.
    }
    \label{fig:mob}
\end{figure}

%\begin{align}
%    \dot phi + \alpha \dot h = \tilde K h'' - B_a, \\
%    \alpha \dot \phi - \dot h = \tilde K \phi'' - \frac{N_n}{2} \sin[2(\phi-\chi)]
%\end{align}
For solving these equations numerically, we form the complex quantity $z = h - i\phi$. Measuring length in units of $D$, time in units of $\frac{1}{\gamma \mu_0 M_s}$ and magnetic field in units of $\mu_0 M_s$, %In those units, 
Eqs. (\ref{eq:7}--\ref{eq:8}) are equivalent to
%\begin{align}
    $(\alpha + i)\dot z = \tilde K z'' - B_a + i\frac{N_n}{2}\sin[2(\phi - \chi)]$, %\label{eq:z}
%\end{align}
where $\tilde K = \frac{2 K_u - \mu_0 M_s^2}{\mu_0 M_s^2}$. 
We employ periodic boundary conditions (PBCs) along $x$ and solve this equation on a GPU by treating the linear part implicitly and the nonlinear part explicitly:
Replacing the derivatives with first-order finite differences, we obtain the semi-implicit numerical equation %of the form 
$T_{ij}z_j(t+\Delta t) = N(t)$,  
%\begin{equation}
%    T_{ij} z_j(t+\Delta t) = -B_a - %\frac{N_n}{2}\sin[2(\phi_i(t) - %\chi_i(t))],
%\end{equation}
where $N(t)$ contains the nonlinear terms, %of Eq.\ (\ref{eq:z}), 
and the matrix $T_{ij}$ is tridiagonal except at the boundaries, where the PBCs give off-diagonal contributions, which can be perturbed away by using the Sherman-Morrison formula \cite{perturbed_tridiag}. This reduces the implicit problem to a tridiagonal linear system, which can be solved using the \verb|cusparseZgtsv2_nopivot| function from CuSparse \cite{cusparse}. 

We choose parameters corresponding to a $0.5~\si{\nano\meter}$ thick Co film within a Pt/Co/Pt multilayer~\cite{metaxas2007creep}, i.e., $K_u = 8.4\cdot 10^5 ~\si{\joule\per\meter^3}$, $A_\ex = 1.4\cdot 10^{-11}~\si{\joule\per\meter}$, $M_s = 9.1\cdot 10^5$ A/m, and $\alpha = 0.27$. We set $\xi = 20~\si{\nano\meter} \approx 3 D$, discretise the DW using a resolution of $6 ~\si{\nano\meter} \approx D$ along $x$, and consider system
sizes from $L\approx 16~\si{\micro\meter}$ up to $L\approx 262~\si{\micro\meter}$.

We start by considering the disorder-dependent steady-state DW velocity $V(B_\ext,\sigma)$. 
For each $\sigma$, an initially flat and uniform (constant $\phi$ along the DW) DW is first let to relax at $B_\ext = 0$ until a static configuration is reached. $B_\ext$ is then increased in steps of $0.1~\si{\milli\tesla}$, evolving for 9.9 $\si{\micro\second}$ at each $B_\ext$-value. The steady-state $V(B_\ext)$ shown in Fig.~\ref{fig:mob}a) 
for different $\sigma$ is
the time-average over the second half of the simulation time. For $\sigma = 0$, $V(B_\ext)$ exhibits a linear increase with $B_\ext$ up to a Walker field $B_\mathrm{W} \approx 2.7~\si{\milli\tesla}$ (in excellent agreement with both the prediction $B_\mathrm{W} = \frac{\alpha}{2}\mu_0 M_\mathrm{s}N_n \approx 2.6~\si{\milli\tesla}$ using $N_n = \frac{\Delta}{\pi D}\ln 2$~\cite{skaugen2019analytical}, and micromagnetic simulations~\cite{herranen2019barkhausen}), at which point $V$ abruptly drops due to the onset of precession of $\phi$. We note that due to weak numerical noise in our implementation, some BLs are present even for $\sigma = 0$, and hence the $\sigma = 0$ curve shown in Fig.~\ref{fig:mob}a) should be interpreted as an "infinitesimal disorder" case.

A finite $\sigma$ results in a non-zero disorder-dependent depinning field $B_c$ where a depinning phase transition takes place.
Above this transition, the velocity curve takes on characteristics of both the Walker breakdown-like effect due to BL nucleation and a sharp increase of $V$ as $B_\ext$ is increased above $B_c$. At very small $\sigma$, BLs are nucleated in large numbers only when $B_\ext$ approaches $B_\mathrm{W}$,
giving rise to a rounded peak in $V(B_\ext)$. As $\sigma$ increases, BLs are nucleated more readily at $B_\ext$-values closer to $B_c$ [insets of Fig.~\ref{fig:mob}a) show example DW configurations for two $\sigma$-values], and BLs are increasingly present also in the initial relaxed state. This causes $V(B_\ext)$ to increase monotonically with $B_\ext$, even when $B_c$ is below the zero-$\sigma$ Walker field. 
For $V(B_\ext)$ in the high-field precessional regime ($B_\ext \gg B_\mathrm{W}$), see Supplemental Material~\cite{SM}.
Fig.~\ref{fig:mob}b) shows a space-time plot of $\phi$ during the dynamics for $B_\ext = B_c \approx 2.22~\si{\milli\tesla}$ at $\sigma = 7~\si{\milli\tesla}$. Notice how the BLs (visible as transitions between $\phi=0$ and $\phi = \pi$ along $x$) nucleate from the initially uniform DW, and subsequently form a dynamic, spatially heterogeneous pattern involving  nucleation, propagation and annihilation of BLs, with the BL density $\rho_{\mathrm{BL}}$ increasing with $t$.

\begin{figure}
    \centering
    \includegraphics[trim=0 0 0 17,width=0.5\textwidth]{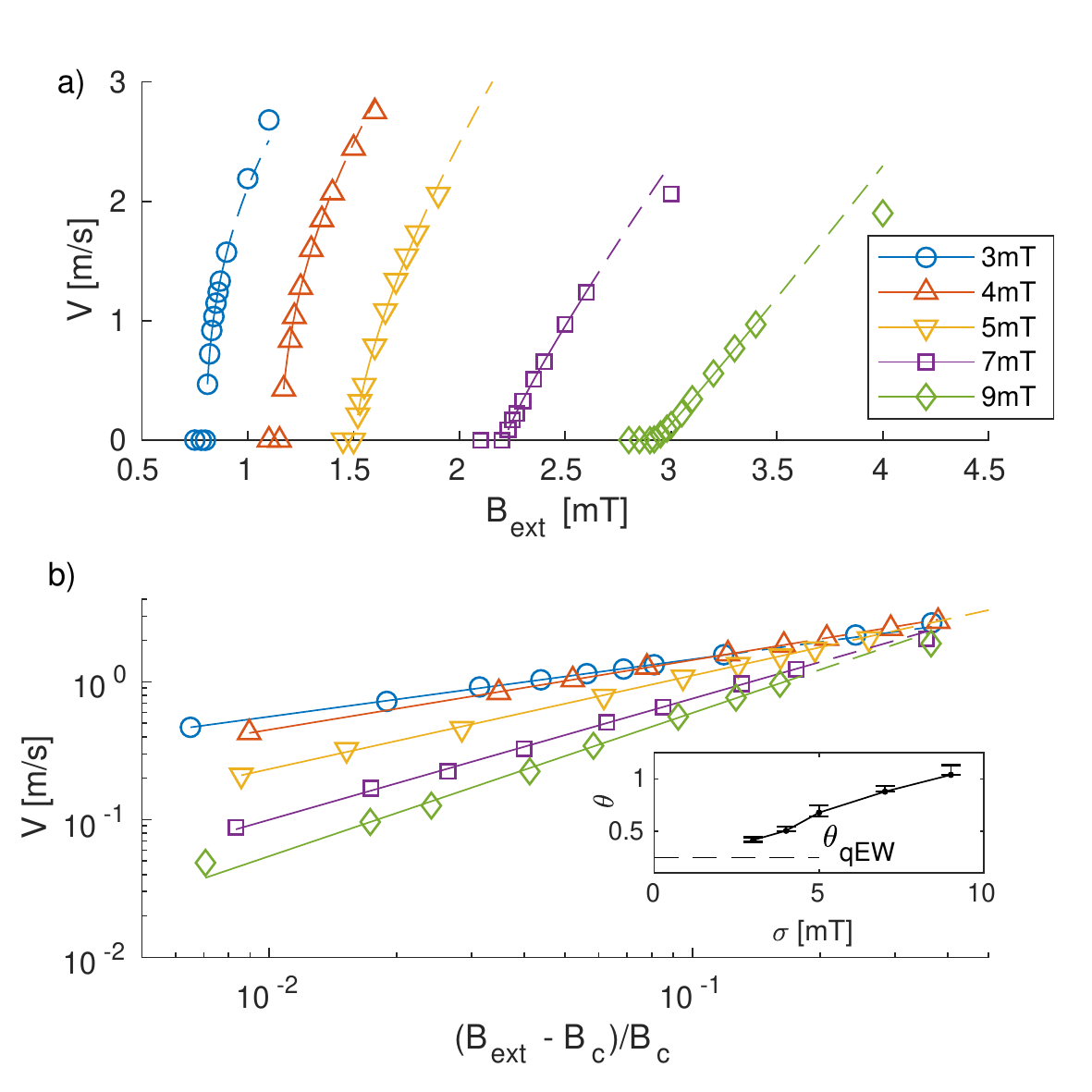}
    \caption{a) Average steady-state DW velocity $V(B_\ext)$ close to $B_c$ for $L\approx 262~\si{\micro \meter}$, for 5 $\sigma$-values (legend). Lines indicate fits of $V = C(B_\ext-B_c)^\theta$. b) Same data shown on a loglog scale 
    as a function of $(B_\ext-B_c)/B_c$. Inset: The
    effective $\sigma$-dependent $\theta$ from the fits, with the 
    dashed horizontal line corresponding to $\theta_{\qEW} = 0.25$.
    }
    \label{fig:depin}
\end{figure}

Close to $B_c$, the system takes on scale-free statistics with large fluctuations, strong finite-size effects, and diverging correlation times. We therefore perform a more careful study in that regime, by averaging over several realizations of the random disorder from a uniform initial condition, using long running times, and varying $L$ from $16$ -- $262~\si{\micro\meter}$. Fig.~\ref{fig:depin} shows the steady-state velocity $V$ close to $B_c$ averaged over 6--50 realizations (with more averaging closer to $B_c$), using 5 different choices of $\sigma$ which lead to values of $B_c$ ranging from well below to well above $B_W$. 
In general, one expects 
$V(B_\ext) \propto (B_\ext-B_c)^{\theta}$; for the qEW equation, $\theta=\theta_\qEW \approx 0.25$ \cite{ferrero2013nonsteady,kim2006depinning}. Fitting a function of this form (lines in Fig.~\ref{fig:depin}), we can 
determine $B_c$ and $\theta$.
Strikingly, as shown in the inset of Fig.~\ref{fig:depin}b), $\theta$ depends on $\sigma$: For small but finite $\sigma$, $\theta$ approaches the qEW value of 0.25, while in the limit of large $\sigma$ it tends to a value close to 1. We note that recent simulations for a specific disorder strength based on the LLG equation of a Heisenberg-like model found %evidence of 
$\theta > \theta_\qEW$ \cite{xiong2018dynamic}. 
%TODO: Move the details from the caption here

\begin{figure}
    \centering
    \includegraphics[width=0.5\textwidth]{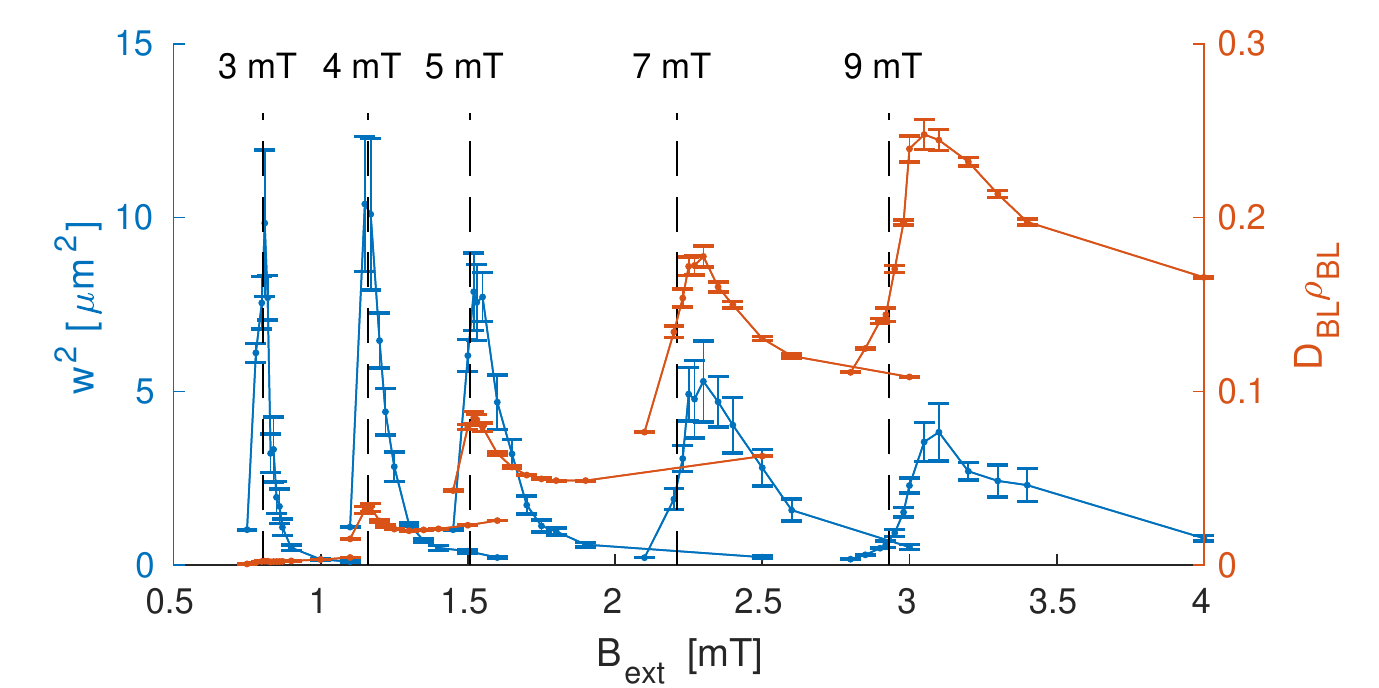} \\
    \caption{Steady-state saturated $w^2(B_\ext)$ for $L\approx 262~\si{\micro\meter}$, for different $\sigma$-values (blue, left axis), and the corresponding BL density $\rho_{BL}(B_\ext)$ in units of the BL width $D_{BL}$ (red, right axis). Vertical dashed lines indicate $B_{c}(\sigma)$.}
    \label{fig:vhnb}
\end{figure}

$\rho_{\mathrm{BL}}$ is found to increase with $\sigma$. 
Fig.~\ref{fig:vhnb} shows that for the smallest $\sigma$ ($\sigma = 3$ mT)
considered, the steady-state $\rho_{\mathrm{BL}}$ is close to zero around $B_\ext = B_c$,
but increases significantly (and exhibits a maximum at or close to $B_\ext = B_c$) with increasing $\sigma$. For large $\sigma$, BL's tend to form heterogeneous arrangements along the DW, with regions of high $\rho_{\mathrm{BL}}$ separated by DW segments essentially free of BLs (see insets of Fig.~\ref{fig:mob}, and the Supplementary Movie \cite{SM}). Concurrently, the squared interface width $w^2 = \fv{(h - \fv{h})^2}$ (averaged over the same number of realizations as for $V$) also displays a maximum close to $B_c$, with the peak value exhibiting
%-- perhaps counter-intuitively -- 
a {\it decrease} with increasing $\sigma$ (Fig.~\ref{fig:vhnb}). 

\begin{figure}
    \centering
    \includegraphics[width=0.5\textwidth]{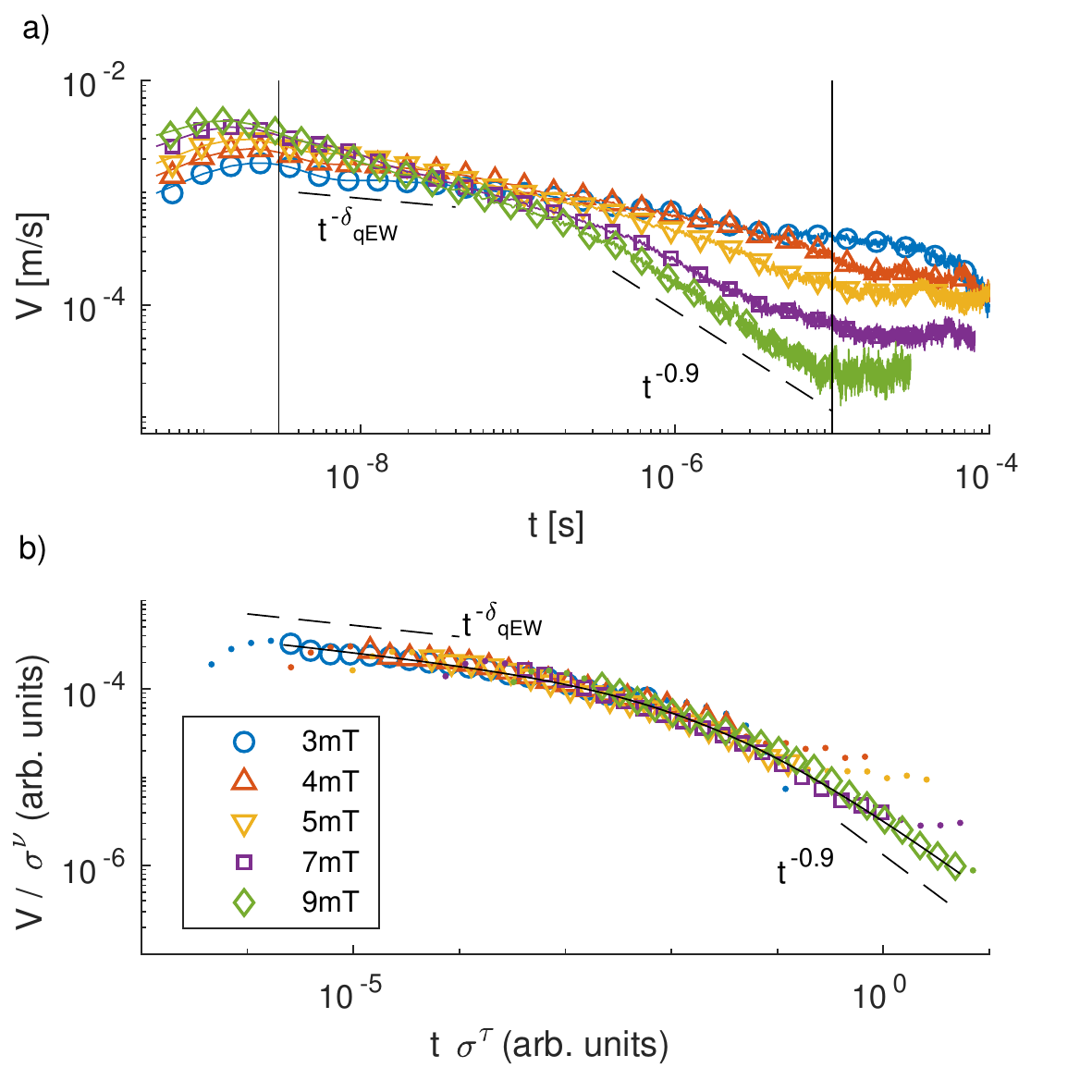}
        \caption{a) Average DW velocity $V(t)$ at $B_\ext = B_c$ for 5  $\sigma$-values [legend in b)] and $L \approx 262~\si{\micro\meter}$. Vertical lines indicate cutoffs applied to the data before collapsing. b) Data collapse of the $V(t,\sigma)$'s
    by rescaling the axes with powers of $\sigma$ ($\tau = 6$ and $\nu = 1.5$).
    Dots indicate data points outside the cutoff lines. Solid line shows a fit of Eq. (\ref{eq:powercross}), while the two dashed lines indicate the asymptotic power laws. 
    }
    \label{fig:collapse}
\end{figure}
\begin{figure}
    \centering
    \includegraphics[width=0.5\textwidth]{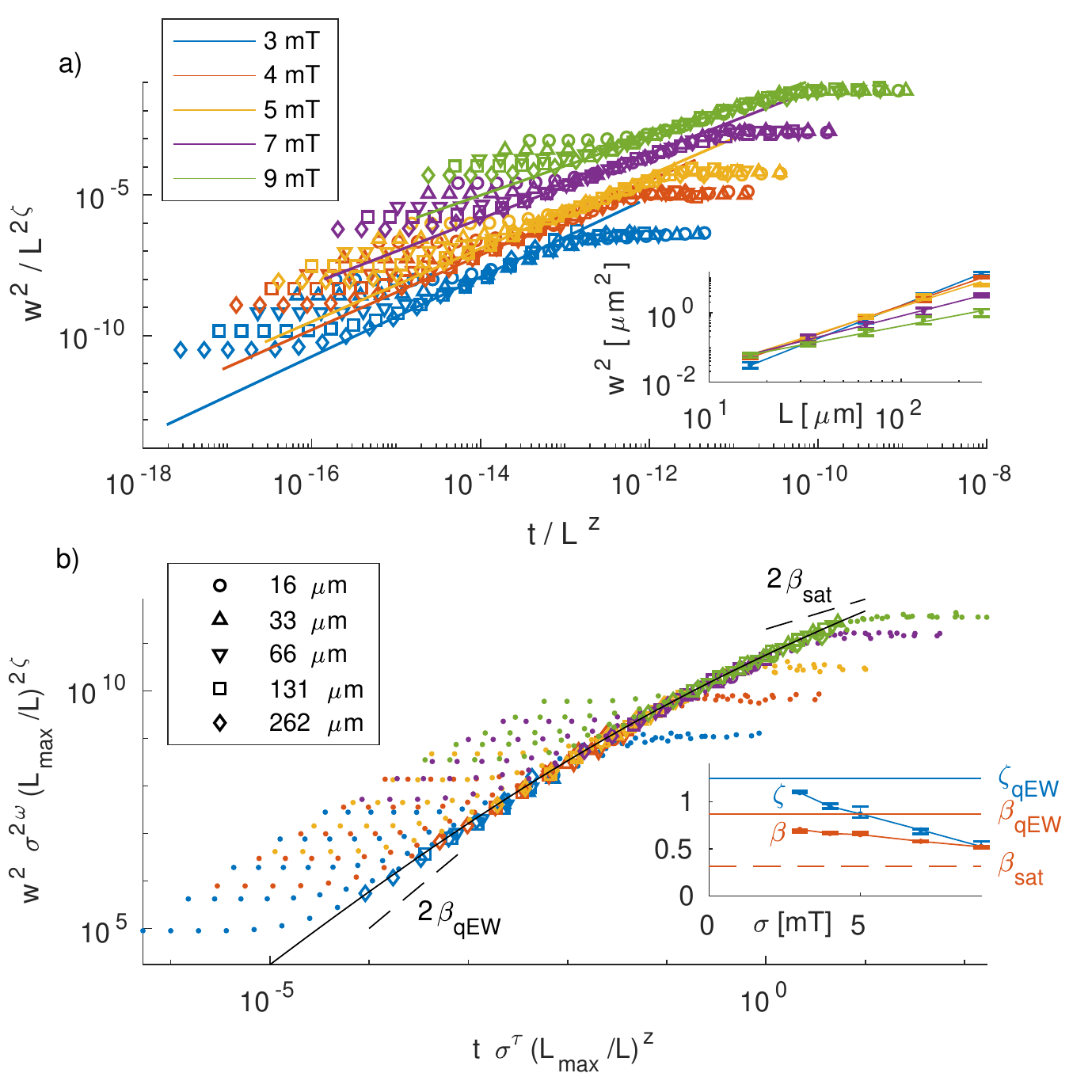}
    \caption{a) Data collapses of $w^2(t,L)$ for each 
    $\sigma$-value (legend).
    Inset: scaling of
    the saturated $w^2$ with $L$ for the different $\sigma$'s. 
    b) Data collapse of the 5 different collapses shown in a), obtained by rescaling with
    powers of $\sigma$ ($\tau = 6$ and $\omega = 7.5$). Solid line shows a fit of the crossover scaling form, and dashed lines indicate the asymptotic power laws. Inset: The $\sigma$-dependent effective exponents $\zeta$ and $\beta$.}
    \label{fig:collapse2}
\end{figure}

Next, we consider the approach to critical pinning by looking at the ensemble-averaged time-dependent velocity $V(t)$, starting from a uniform state at $B_\ext = B_c$, as obtained above from the fits to the steady-state velocities. We expect $V(t)$ to follow $V(t) \propto t^{-\delta}$, with $\delta_\qEW = 0.129$ for the qEW equation~\cite{ferrero2013nonsteady,kim2006depinning}. Fig.~\ref{fig:collapse}a) shows the $V(t)$'s for the five different $\sigma$'s considered (symbols represent logarithmically binned data, plotted on top of the raw averaged velocity signals shown with lines). At the small $\sigma = 3~\si{\milli\tesla}$, we see an early power law strongly resembling the qEW behavior, but at late times $V(t)$ begins a transition to a steeper decay. 
As $\sigma$ is increased, this transition becomes more dramatic and happens at earlier times, until we see a long-$t$ strong-disorder "saturated" value
$\delta=\delta_{\mathrm{sat}}$ close to 0.9 at the highest $\sigma$. This is associated with $\rho_{\mathrm{BL}}$ 
increasing both with $t$ and $\sigma$ (see Figs.~\ref{fig:mob} and~\ref{fig:vhnb}). We note that this results in a $t$ and $\sigma$-dependent DW mobility via  
    $V/B_\ext = D/(\alpha + \frac{\pi^2}{2\alpha}D_\mathrm{BL}\rho_\mathrm{BL})$,
where $D_{BL}$ is the BL width~\cite{malozemoff1972effect}. 

A possible interpretation is a disorder-dependent crossover timescale $t_c(\sigma)$ between two different power law regimes. To test this, we rescale the time axis by $\sigma^\tau$ and the velocity by $\sigma^\nu$.
Fig.~\ref{fig:collapse}b) shows the logarithmically binned data, rescaled with $\tau = 6$ and $\nu = 1.5$, resulting in a good data collapse of the central parts of the velocity signals (large symbols). We then fit the resulting master curve with the crossover scaling form \cite{laurson2014universality}
\begin{equation}
    V(t,B_\ext = B_c) = C t^{-\delta_{\qEW}}
    \left[1 + \left(\frac{t}{t_c}\right)^{k(\delta_{\mathrm{sat}} - \delta_{\qEW})} \right]^{-1/k},
    \label{eq:powercross}
\end{equation}
shown as a solid black line in Fig.~\ref{fig:collapse}b); the asymptotic power laws $V(t\ll t_c) \propto t^{-\delta_\qEW}$ and $V(t\gg t_c) \propto t^{-\delta_{\mathrm{sat}}}$ are indicated as dashed lines. $k$ controls the sharpness of the crossover; our fit gives $k \approx 0.48$, consistent with the relatively slow crossover.

Having determined the $\sigma$-dependent $\theta$ and $\delta$, we finally study the roughness of the DW at $B_\ext = B_c$. 
$w$ is expected to follow the scalings $w \propto t^{\beta}$ for $t \ll t^*$ and $ w \propto L^{\zeta}$ for $t \gg t^*$, where $t^* \propto L^z$ with $z = \zeta/\beta$ \cite{vicsek1984dynamic}, with $\zeta$ and $z$ the roughness and dynamic exponents, respectively.
Fig.~\ref{fig:collapse2}a) shows the data collapses according to these scalings separately for the 5 $\sigma$-values (saturated $w^2$ 
shown in the inset), allowing
us to obtain estimates of $\sigma$-dependent $\beta$ and $\zeta$ 
[inset of Fig.~\ref{fig:collapse2}b)]. %Again, 
In the limit of small but finite $\sigma$, the exponents tend towards the qEW values. To estimate the asymptotic strong $\sigma$ $\beta$-exponent, we first collapse the 5 disorder-specific data collapses in Fig.~\ref{fig:collapse2}a) 
by %again 
rescaling the data with powers of $\sigma$, 
resulting in a good data collapse for the middle parts of the data [large symbols in Fig.~\ref{fig:collapse2}b)]. Fitting the master curve with a crossover scaling form similar to Eq.~(\ref{eq:powercross}) reveals %again 
a slow crossover ($k\approx 0.2$) from $\beta_\mathrm{qEW} \approx 0.87$ \cite{kim2006depinning} for early times/weak disorder to an asymptotic long time/strong disorder exponent $\beta_\mathrm{sat} \approx 0.32$. Concurrently, we find an effective $\zeta$ decreasing from $\zeta_{\mathrm{qEW}}\approx 1.25$ as $\sigma$ is increased [inset of Fig.~\ref{fig:collapse2}b)].

To conclude, our results reveal the unusual disorder-dependent nature of criticality at the depinning transition of thin film DWs. These features are related to localized reduction of DW mobility for strong disorder due to proliferation of BLs within moving parts of the DW, damping the roughness growth. This indicates that simple elastic line models are unable to properly describe depinning dynamics of DWs. Experimental studies verifying these results are needed, and are likely to be challenging due to the need to reach very low temperatures \cite{albornoz2021universal,gorchon2014pinning} (to minimize thermal rounding \cite{bustingorry2012thermal}), and to control the disorder. %of the thin film samples. 
It would %also be of interest 
be interesting to extend our study to 3D systems with 2D DWs with internal degrees of freedom~\cite{herranen2017blochline}, to consider effects due to a finite $T$ \cite{jin2018numerical}, as well as the interplay of $\xi$ with the DW and BL widths. 
%Finally, we point out that 
Our model should find applications in modelling DW dynamics in a wide range of contexts where DW velocities are not so high that spin wave emission from the moving DW~\cite{yoshimura2016soliton,voto2016effects,voto2016disorder} (not captured by any model which limits the description to the degrees of freedom of the DW) becomes important, including creep motion of DWs~\cite{metaxas2007creep} and Barkhausen noise~\cite{durin2006science}. Finally, extensions to bubble geometry would be useful, e.g., for studying effects due to the Dzyaloshinskii-Moriya interaction %on DW dynamics
\cite{diez2019enhancement, vandermeulen2018comparison}.  

\bibliographystyle{apsrev4-2}
%\bibliography{ref}

\end{document}